\documentclass[prb,amsmath,amssymb,superscriptaddress,twocolumn]{revtex4}
\usepackage{graphicx, slashed, color}

\newcommand{\eps}{\epsilon}

\DeclareMathAlphabet{\mathpzc}{OT1}{pzc}{m}{it} 
\begin{document}
\title{Quantum superconducting criticality in graphene and  topological insulators }

\author{Bitan Roy}
\affiliation{National High Magnetic Field Laboratory, Florida State University, Tallahassee, Florida 32306, USA}
\author{Vladimir Juri\v ci\' c}
\affiliation{Instituut-Lorentz for Theoretical Physics, Universiteit Leiden, P.O. Box 9506, 2300 RA Leiden, The Netherlands}
\author{Igor F. Herbut}
\affiliation{ Department of Physics, Simon Fraser University, Burnaby, British Columbia, Canada V5A 1S6}
\affiliation{ Max-Planck-Institut f\"ur Physik Komplexer Systeme, N\"othnitzer Str. 38, 01187 Dresden, Germany}
\date{\today}
\begin{abstract}
 The field theory of the semimetal-superconductor quantum phase transition for graphene and surface states of topological insulators is presented. The Lagrangian possesses the global $U(1)$ symmetry, with the self-interacting complex bosonic order-parameter and the massless Dirac fermions coupled through a Yukawa term. The same theory also governs the quantum critical behavior of graphene near the transition towards the bond-density-wave (Kekule) insulator. The  local $U(1)$ gauged version of the theory which describes the quantum semimetal-superconductor transition in the ultimate critical regime is also considered. Due to the Yukawa coupling the transitions are found to be always continuous, both with and without the fluctuating gauge field. The critical behavior is addressed within the dimensional regularization near four space-time dimensions, and the calculation of various universal quantities, including critical exponents and the universal mass-ratio, is reported.

\end{abstract}


\maketitle

Dirac quasiparticles represent low-energy excitations in various low-dimensional condensed-matter systems, such as graphene and topological insulators (TIs). In graphene, pseudorelativistic Dirac quasiparticles emerge from  hopping of the electrons on the underlying honeycomb lattice,\cite{wallace,semenoff} while on the surface of a strong (crystalline) TI, they result from an odd (even) number of band inversions in the bulk  of the system.\cite{fu-kane-3D} In all these cases, the linearly dispersing Dirac quasiparticles give rise to a semimetallic ground state, stable against weak electron-electron  interactions.\cite{herbut-physics}

When the repulsive interactions are sufficiently strong, however, a plethora of insulating phases can in principle be realized in graphene.\cite{herbut-physics,  HJR, chamon} Furthermore, Dirac fermions in graphene can also condense into \emph{four} different gapped superconducting states, if the net interaction acquires an attractive component.\cite{chamon, igor-bitan-SC} The simplest of them, which will be the subject of the present study, is the uniform, spin-singlet $s$-wave pairing, favored by a sufficiently strong on-site attractive interaction.\cite{zhao} Two of the remaining pairing gaps are spatially inhomogeneous spin triplets, which break the translational symmetry of the honeycomb lattice into Kekule patterns.\cite{igor-bitan-SC} They are favored by the sufficiently strong nearest-neighbor attraction. Finally, yet another triplet pairing with an $f$-wave symmetry can be stabilized by a strong second-neighbor attraction.\cite{honerkamp, otherSCgraphene} On the other hand, due to their reduced number of fermionic components, the massless Dirac fermions residing on the surface of TIs with a single surface Dirac cone can acquire a superconducting gap only by pairing to an $s$-wave superconducting state. Possible inhomogeneous \cite{igor-bitan-SC} and chiral $d+i d$ superconducting states in doped graphene,\cite{chubukov, thomale} as well as proposed realizations of Majorana fermions in graphene and TIs, \cite{fu-kane-Majorana,linder,ghaemi} make the study of superconducting instabilities of Dirac fermions in low-dimensional condensed-matter systems theoretically and experimentally interesting and timely. \cite{metzner,grover-vishwanath,goldhaber,molenkamp}

The \emph{bosonic} order parameters (OPs), characterizing both the insulating and the superconducting states, are composite objects of Dirac fermions, and may exhibit different  symmetries. Besides the usual self-interaction, the OPs here are also coupled to the massless Dirac fermions via the Yukawa term.\cite{HJV-PRB} Previously, we studied the Ising and Heisenberg universality classes of the transition into charge-density-wave and spin-density-wave insulators, respectively. The Cooper pairs are of course {\it charged}, and consequently the corresponding field theory possesses a global $U(1)$ symmetry. We therefore here develop the $U(1)$-symmetric field-theoretical description of the quantum semimetal-superconductor transition in graphene and surfaces of TIs. All the coupling constants in this effective theory are \emph{marginal} in $d=4$ space-time dimensions, which enables us to perform the $\epsilon-$ expansion with $\epsilon=4-d$ to address its critical behavior. We find that the semimetal-superconductor transition is always continuous, and we compute the critical exponents associated with the transition; in particular, the correlation-length exponent and the anomalous dimensions for both the OPs and the fermion fields are obtained. Besides describing the semimetal-superconductor quantum phase transition, the $U(1)$ field theory should also pertain to the quantum phase transition from the semimetallic into an insulating state with the dynamically generated Kekule mass, which breaks the translation symmetry of the lattice.\cite{hou-chamon-mudry} Motivated by this physical problem, as well as by the theoretical possibility of a fully Lorentz invariant semimetal-superconducting transition in these Dirac systems, we extend our theory to include a fluctuating $U(1)$ gauge field, that would  describe the coupling of the electromagnetic field to the bosonic OP and the Dirac fermions near the critical point.\cite{igor-book} The Lorentz-symmetric critical point describing the continuous transition in this theory is {\it charged} and also turns out to be stable for any physical number of fermion flavors (that is, for $N_f \geq 0.142$).

To set up the problem, we first consider the pairing of the the gapless excitations in graphene, around the two inequivalent Dirac points, at $\pm {\bf K}$, described by an eight-component Dirac-Nambu spinor, $\Psi^\dagger(k)=[\Psi_+^\dagger(k),\Psi_-^\dagger(k)]$, where $\Psi_\sigma^\dagger(k)= \left[ u_\sigma^\dagger(k),v_\sigma^\dagger(k),\sigma u_{-\sigma}(-k),\sigma v_{-\sigma}(-k) \right]$.\cite{Herbut-topology} Here, ${\bf K}=(1,1/\sqrt{3}) (2 \pi/a \sqrt{3})$, with $a$ being the lattice constant. $k\equiv(\omega,{\bf k})$ is the three-momentum and ${\bf k}={\bf K}+{\bf q}$, $|{\bf q}| \ll |{\bf K}|$. $\sigma=\pm$ is the spin projection along the $z$ axis. $u_\sigma$ and $v_\sigma$ are the Grassmanian fields on two sublattices. The free Dirac Lagrangian in this representation assumes the relativistically invariant form $L_f^0=i {\bar \Psi}(x)\sigma_0 \otimes\gamma_\mu\partial_\mu\Psi(x)$, where $\Psi(x)=\int d^3k e^{ikx}\Psi(k)$, $\mu=0,1,2$, and $x\equiv(\tau,{\bf r})$ with $\tau$ as the imaginary time and summation over the repeated indices assumed. The $\gamma$ matrices are defined as $\gamma_0=\sigma_3 \otimes\sigma_3, \gamma_1=\sigma_0 \otimes\sigma_2$, $\gamma_2=\sigma_0 \otimes\sigma_1$, $\gamma_3=\sigma_1\otimes\sigma_3$, and $\gamma_5=\sigma_2\otimes\sigma_3$, where $\{ \sigma_0, {\boldsymbol{\sigma}} \}$ forms the Pauli basis for two-dimensional matrices, and we take ${\bar\Psi}\equiv\Psi^\dagger\gamma_0$, as usual. The $s$-wave superconducting OP reads
\begin{equation}
\Phi(x)=\langle\Psi^\dagger(x)\sigma_0 \otimes (i\gamma_0\gamma_3\cos\varphi+i\gamma_0\gamma_5\sin\varphi)\Psi(x)\rangle,
\label{bosonicOP}
\end{equation}
with $\varphi$ as the {\it phase} of the superconducting OP. The OP anticommutes with the particle number operator $N=\sigma_0 \otimes i\gamma_3\gamma_5$ and commutes with all the three generators of the spin rotations, ${\bf S}=\boldsymbol{\sigma} \otimes I_4$, and hence represents a spin singlet. Moreover, it is even under the exchange of the sublattices, or of the Dirac points.

In terms of Nambu's (particle-hole doubled) spinor basis, $\Psi^{\dagger}= (c^\dagger_{\vec k \uparrow},c^\dagger_{\vec k \downarrow}, c_{-\vec{k} \downarrow},-c_{-\vec{k},\uparrow})$, the Lagrangian for the gapless surface states of TIs also adopts the relativistic form $L_f^0=i {\bar \Psi}(x)\gamma_\mu\partial_\mu\Psi(x)$, with the $\gamma$ matrices defined as $\gamma_0=\sigma_3 \otimes \sigma_3, \gamma_1=- \sigma_0 \otimes \sigma_1, \gamma_2=\sigma_0 \otimes \sigma_3, \gamma_3=\sigma_2 \otimes \sigma_3$, and $\gamma_5=\sigma_1 \otimes \sigma_3$. The $s$-wave superconducting OP and the number operator for the surface states of TIs assume a form identical to that for graphene, only without the $\sigma_0$ in the first block. Due to Nambu's particle-hole doubling, the true number of fermionic degrees of freedom on the surface of a TI is a {\it quarter} of the one in graphene, however. In this special case the critical theory acquires the supersymmetry, and the one-loop $\epsilon$ expansion is known to be exact.\cite{sslee}

Next we wish to study the quantum phase transition from the semimetallic into the $s$-wave superconducting phase. Since we want to formulate an $\epsilon (=4-d)$-expansion near \emph{four} space-time dimensions, we need first to define a spinor basis in which the theory can be formally extended from the physical three dimensions to four space-time dimensions. We therefore rotate the spinor $\Psi\rightarrow U\Psi$, where, in graphene, $U=\exp\left[ {i\frac{\pi}{4}\sigma_0 \otimes\gamma_3} \right]$. After this unitary transformation, the $s$-wave OP reads
\begin{equation}
\Phi(x)=\langle\Psi^\dagger(x)(\sigma_0 \otimes \gamma_0 \cos\varphi +\sigma_0\otimes i\gamma_0\gamma_5 \sin\varphi)\Psi(x)\rangle,
\label{s-wave4d}
\end{equation}
while leaving the relativistically invariant free Dirac Lagrangian, $L^0_f$, unchanged. The number operator is then ${\hat N}=\sigma_0 \otimes\gamma_5$. Similarly, in a TI, the analogous transformation is performed by choosing the simpler $U=\exp\left[ {i\frac{\pi}{4}\gamma_3} \right]$.

For generality, we consider the $U(1)$ gauge theory for $N_f$ flavors of four-component Dirac fermions coupled to the bosonic OP with $N_b$ complex components via the Yukawa coupling in the presence of a fluctuating gauge field,  with the complete Lagrangian
$L=L_f+L_b+L_{bf}+L_{EM}$. The coupling of the fermions to the $U(1)$ gauge field reads
\begin{equation}
L_f={\bar \Psi}(x)\gamma_\mu(\partial_\mu-ie\gamma_5A_\mu)\Psi(x).
\end{equation}
The matrix  $\gamma_5$ appearing in the minimal coupling is then the number operator, and $e$ is the $U(1)$ charge. The coupling of the OP to the massless fermions has the Yukawa form
\begin{equation}
L_{bf}=g[({\rm Re}\Phi){\bar \Psi}\Psi+({\rm Im}\Phi){\bar \Psi}i\gamma_5\Psi].
\end{equation}
On the other hand, the dynamics of the OP coupled to the $U(1)$ gauge field can be described by the standard Ginzburg-Landau Lagrangian
\begin{equation}
L_b=|(\partial_\mu+2ieA_\mu)\Phi|^2+m^2|\Phi|^2+\frac{\lambda}{2}|\Phi|^4,
\end{equation}
where $m^2$ is the tuning parameter of the transition. The $U(1)$ gauge field is described by the usual Maxwell Lagrangian
\begin{equation}
L_{EM}=\frac{1}{4}F_{\mu\nu}F_{\mu\nu},
\end{equation}
with $F_{\mu\nu}=\partial_{\mu}A_\nu-\partial_\nu A_\mu$. We use the transverse (Landau) gauge $\partial_\mu A_\mu=0$, in which the general gauge-invariant OP becomes local.\cite{igor-book} The above theory is constructed to be invariant under the following local $U(1)$ gauge transformation $\Psi \rightarrow e^{ie\theta\gamma_5}\Psi, \: \Phi \rightarrow e^{-2ie\theta}\Phi, \: A_\mu\rightarrow A_\mu+\partial_\mu\theta$.

Without the gauge field $(e=0)$, the above field theory for $N_f=2$ and $N_b=1$ also governs the critical behavior of graphene close to the transition to the spin-singlet Kekule insulator. However, in the latter case, the Dirac spinor needs to be redefined as $\Psi=\left[ \Psi_+, \Psi_-\right]^\top$, where $\Psi_\sigma=\left[u_\sigma({\bf K}+{\bf q}),v_\sigma({\bf K}+{\bf q}),u_\sigma(-{\bf K}+{\bf q}),v_\sigma(-{\bf K}+{\bf q}) \right]$, with $\sigma= \pm$ as the spin projections and with the frequency label suppressed.\cite{gamma-insulator} We have set the Fermi velocity and the velocity of the bosonic excitations to be equal, since a weak anisotropy in the velocities is irrelevant.\cite{HJV-PRB} The local ($e\neq 0$)  $U(1)$ gauge theory describes the ultimate critical behavior at the superconducting transition, at which {\it all} the velocities in the theory are equal to the velocity of light. In graphene and on the surface of TIs, however, such a fixed point is experimentally inaccessible, since the bare Fermi velocities are $\sim 10^6$ m/s.

 Next, we proceed with the analysis of the $U(1)$-symmetric Yukawa field theory. The couplings $\lambda$, $e$, and $g$ are all dimensionless in (3+1) space-time dimensions, suggesting the $\epsilon$ expansion about $d+1 =4$ as a tool of choice for the study of the quantum critical behavior. Define then the action $S_{\rm ren}=\int d\tau\int d^d x L_{\rm ren}$, where the renormalized Lagrangian is
\begin{eqnarray}
L_{\rm ren}&=&Z_\Psi L_f+ Z_\Phi|(\partial_\mu+2ieA_\mu)\Phi|^2+Z_m m^2|\Phi|^2 \nonumber \\
&+& Z_\lambda\frac{\lambda}{2}|\Phi|^4 +Z_g L_{bf} + Z_A L_{EM}.
\end{eqnarray}
The computation of the self-energy diagrams for the fermions, the order parameter, and the gauge field using a minimal-subtraction scheme then yields the renormalization constants to the one-loop order
\begin{equation} \label{z-phi-psi}
Z_\Psi=1-\frac{1}{2}g^2\frac{1}{\epsilon}, \quad Z_\Phi=1-g^2N_f\frac{1}{\epsilon}+12e^2\frac{1}{\epsilon},
\end{equation}
\begin{equation}
Z_A=1-e^2\frac{4}{3}(N_f+N_b)\frac{1}{\epsilon},\label{Z-A}
\end{equation}
where $\epsilon=4-d$ and the dimensionless couplings $Q=\{e^2,g^2,\lambda\}$ are rescaled as $Q S_d/(2\pi)^d \rightarrow Q$, with $S_d=2\pi^{d/2}/\Gamma(d/2)$. (See Supplementary Material). The computation of the vertex diagrams to the same order gives the following renormalization conditions for the coupling constants:
\begin{equation}
Z_\Psi Z_\Phi^{1/2}g_0\mu^{-\epsilon/2}+3e^2g\frac{1}{\eps}=g,\label{RC-g}
\end{equation}
\begin{equation}
Z_\Phi^2\lambda_0\mu^{-\epsilon}-\lambda^2(N_b+4)\frac{1}{\epsilon}-96e^4\frac{1}{\epsilon}
+2g^4N_f\frac{1}{\epsilon}=\lambda.
\label{RC-lambda}
\end{equation}
The renormalization of the tuning parameter $(m^2)$ can be extracted from the self-energy diagrams of the OP, leading to
\begin{equation}\label{corr-length-exp}
Z_\Phi m_0^2\mu^{-\epsilon}-\lambda(N_b+1)\frac{1}{\epsilon}m^2=m^2.
\end{equation}
Here, the couplings with subscript ``0'' are the bare couplings, the ones without the subscript are the renormalized couplings, and $\mu$ is the renormalization scale. Dimensional regularization explicitly preserves gauge invariance of the theory implying $\mu^{-\epsilon}Z_Ae_0^2=e^2$, to {\it any} order. \cite{igor-zlatko, igor-book} In conjunction with this identity and Eq.\  (\ref{Z-A}), one can write the (ultraviolet) $\beta$ function of the charge as
\begin{equation}
\beta_{e^2}\equiv\frac{d e^2}{d\ln\mu}=-\epsilon e^2+\frac{4}{3}(N_f+N_b)e^4.
\end{equation}
The renormalization group flow of the remaining two couplings can be obtained from Eqs. (\ref{RC-g}) and (\ref{RC-lambda}):
\begin{eqnarray}
\beta_{g^2}&=&-\epsilon g^2+(N_f+1)g^4-18e^2g^2,\\
\label{beta-g}
\beta_\lambda&=&-\epsilon\lambda+2N_fg^2 (\lambda -g^2)-24e^2 (\lambda -4 e^2)\nonumber\\
&+&(N_b+4)\lambda^2.
\label{beta-lambda}
\end{eqnarray}
The above $\beta$ functions, besides the trivial, yield the following {\it neutral} ($e =0$) fixed points. (1) the Wilson-Fisher fixed point at  $g_*^2=0$ and $\lambda_*=\epsilon/(N_b+4) $. (2) The neutral Gross-Neveu fixed point: $(g_*^2,\lambda_*)=\left(\epsilon/X, (a+b)\epsilon\right),$ where $a=(1-N_f)/2 X W, b=\sqrt{(N_f-1)^2+ 8 N_f W}/2 X W$, with $X=N_f+1$ and $W= N_b+4$. Ignoring the gauge coupling $e$ for the moment, this fixed point is \emph{critical}, and it controls the transition towards the $s$-wave superconducting state or into the spin-singlet Kekule insulator. Weak charge $e^2$ is, however, a relevant coupling at this critical point. (3) The bicritical point in the $e^2=0$ plane: $\left(g_*^2,\lambda \right)=\left(\epsilon/X, (a-b) \epsilon \right)$, located in the unphysical region ($\lambda <0$) of the $\Phi^4$-interaction. Therefore, our one-loop results suggest that the semimetal-superconducting transition is of the second order in the absence of the fluctuating gauge field. The result is qualitatively similar to the insulating Ising and Heisenberg universality classes. \cite{HJV-PRB} The correlation-length exponent ($\nu$) can readily be determined from Eqs.\ (\ref{z-phi-psi}) and (\ref{corr-length-exp}), yielding
\begin{equation}\label{correlationlength}
\nu=\frac{1}{2}+\frac{1}{4}(N_b+1)\lambda_*-3e^2_*+\frac{N_f}{4}g_*^2,
\end{equation}
with $e^2_*=0$ and $(g^2_*,\lambda_*)$ corresponding to the neutral Gross-Neveu critical point. Since the Lorentz-symmetry-breaking perturbations are irrelevant near the critical point,\cite{HJV-PRB} the dynamical critical exponent is $z=1$, and the Fermi velocity ($v_F$) is non-critical. Near the neutral critical point both the OPs and the fermion fields acquire nontrivial anomalous dimensions, which read, respectively, 
\begin{equation} \label{anomalousdimension}
\eta_b=\left( g^2_* \; N_f-12 e^2_* \right) \epsilon+{\cal O}(\epsilon^2), \:
\eta_f=\frac{g^2_*}{2}\epsilon+{\cal O}(\epsilon^2).
\end{equation}

The residue of the quasiparticle pole of the fermions $Z\sim m^{z\nu\eta_f}\sim m^{\epsilon/4X}$, and they cease to exist as sharp excitations at the critical point. Moreover, as the system approaches the critical point from the superconducting side both the mass of the superconducting OP and the fermion mass vanish with the universal ratio
\begin{equation}\label{massratio}
\frac{m_b^2}{m_f^2}=\frac{2\lambda_*}{g_*^2}.
\end{equation}
In order to extract the critical exponents and amplitudes for graphene one needs to substitute $N_f=2$ and $N_b =1$, while for the surface states of TIs one should use $N_f=1/2$ and $N_b = 1$. In the latter case, we obtain $\eta_b=\eta_f=\epsilon/3$ and $\nu=1/2+\epsilon/4$ in agreement with Ref.\ \onlinecite{sslee}.

In the fully gauged theory with $e\neq 0$, the {\it charged} Wilson-Fisher fixed points are at $\left( e^2_*,g^2_*,\lambda^\pm_* \right)=\left( \frac{3}{4 Y},0,\frac{18+Y \pm \sqrt{(18+Y)^2-216 W}}{2 Y W}\right) \epsilon$, where $Y=N_f+N_b$. On the other hand, the previously discussed neutral Gross-Neveu fixed point is unstable in the charge direction, and a pair of charged fixed points is located at
\begin{equation}\label{chargedcritical}
\left( e_*^2,g_*^2,\lambda_*^\pm \right) = \left(\frac{3}{4 Y},\frac{27+2Y}{2 X Y},\frac{\left[\Delta_1\pm\sqrt{\Delta_1^2+\Delta_2}\right]}{2 X^2 Y^2 W} \right) \epsilon,
\end{equation}
where $\Delta_1= X Y \left[ X Y+ 18 X-N_f (27+2 Y) \right]$ and $\Delta_2=-4 W X^2 Y^2 \left[ 54 X^2-2N_f (13.5+ Y)^2 \right]$. However, only the fixed point with $\lambda_*^+>0$ is stable in the critical plane ($m^2=0$). This fixed point is therefore critical and controls the behavior in the vicinity of the quantum phase transition  in the full Lorentz-invariant $U(1)$ gauge theory. Furthermore, for any physical number of flavors this critical point describes the second-order phase transition, since the quantity $\Delta_1^2+\Delta_2$  is {\it positive} for any $N_f \geq 0.142$. However, since all the velocities in this theory are set to be equal to the velocity of light, this critical point may be reached only in the deep infrared regime.\cite{sachdev-vojta} The other fixed point at $(e^2_*,g_*,\lambda^{-}_*)$ lies in the unphysical region ($\lambda<0$) of the $\Phi^4$ interaction for any $N_f$. On the other hand, when $N_f=0$, we obtain the standard one-loop result for the critical number of the complex components of the OP above which the superconductor transition is of the second order, $N_b^{crit}\simeq182.952$, and the transition is controlled by the charged Wilson-Fisher fixed point, with $g^2_*=0$.\cite{igor-zlatko, igor-book} It is worth observing that without the Yukawa interaction, $N_b^{crit}$ reduces to $3.47$ if one takes into account only the coupling of fluctuating gauge fields with the massless Dirac fermions.\cite{Nogueira} The Yukawa coupling therefore appears to be crucial for the stabilization of the criticality in the theory and for the suppression of the possible discontinuous transition, which occurs in related theories. \cite{igor-book}

The superconducting coherence length ($\xi$) diverges as $\xi \sim m^{-\nu}$, and the correlation length exponent ($\nu$) can be computed readily from Eq. (\ref{correlationlength}). The boson and the fermion fields in the vicinity of this charged critical point acquire anomalous dimensions, which can be found from Eq. (\ref{anomalousdimension}). One can also compute the flow for the Ginzburg-Landau parameter $\kappa^2=\lambda/(2e^2)$ characterizing the transition:
\begin{eqnarray}
\beta_{\kappa^2}&=&e^2\left[2 W \kappa^4 -2 \bigg\{ \frac{2}{3}Y+12 \bigg\} \kappa^2 +48\right.\nonumber\\
&+&\left.2 N_f \bigg(\frac{g^2}{e^2} \bigg) \bigg\{ \kappa^2-\frac{1}{2} \; \bigg(\frac{g^2}{e^2} \bigg) \bigg\}\right].
\end{eqnarray}
At the \emph{charged} Gross-Neveu critical point, this flow equation has fixed points at $\kappa^2_-<0$ and $\kappa_+^2>0$, for arbitrary $N_b$ and $N_f \geq 0.142$. The residue of the quasiparticle pole vanishes at the charged critical point as $Z\sim m^{\frac{27+2 Y}{8 X Y}\epsilon+{\cal O}(\epsilon^2)}$.

Topological crystalline insulators, such as the recently observed SnTe in Ref.\ \onlinecite{tanaka} and Sn-doped PbTe and PbSe in Ref.\ \onlinecite{pbbasedtopocrys}, host four Dirac cones on the surface amounting to $N_f=4 \times 1/2=2$ species of four-component Dirac fermions.
The possibility of the superconducting transition on the surface of topological crystalline insulators makes our theory relevant for this problem as well. The critical behavior in this case is captured within our theory upon substituting $N_f=2$ and $N_b=1$, same as in graphene.

The optical conductivity in the entire semimetallic phase remains constant and  universal, while it becomes infinite (zero) in the superconducting (Kekule) phase. Right at the quantum critical point it is also expected to be universal, but different from the one in the semi-metallic phase.\cite{HJR}  The universal conductivity at the Gross-Neveu (neutral or charged) critical point is also expected to be different from one found in a pure bosonic theory. \cite{herbut-universalcond} The computation of its value is an interesting problem left for future research.

To summarize, by employing a $U(1)$-symmetric Gross-Neveu-Yukawa theory, we here studied the zero-temperature semimetal-superconductor (Kekule insulator) transition in graphene and on the surface of TIs, and showed that it is continuous for any number of Dirac flavors. The full $U(1)$ gauge theory exhibits a \emph{charged} critical point also for an arbitrary number of Dirac flavors, and may be relevant for the semimetal-superconducting transition in the deep infrared regime.

B.R. wishes to acknowledge the support of National Science Foundation Cooperative Agreement No. DMR-0654118, the State of Florida, and the U.S. Department of Energy. V.J. acknowledges the support of the Netherlands Organization for Scientific Research (NWO). I.F.H. is supported by the NSERC of Canada.

\pagebreak

\onecolumngrid

\begin{center}
{\bf \large Supplementary material of ``Quantum superconducting criticality in graphene and  topological insulators''} \\
{Bitan Roy$^1$, Vladimir Juri\v ci\' c$^2$, and Igor F. Herbut$^3$ }
\end{center}

\begin{center}
{$^1$ National High Magnetic Field Laboratory, Florida State University, Tallahassee, Florida 32306, USA \\
$^2$ Instituut-Lorentz for Theoretical Physics, Universiteit Leiden, P.O. Box 9506, 2300 RA Leiden, The Netherlands \\
$^3$Department of Physics, Simon Fraser University, Burnaby, British Columbia, Canada V5A 1S6 \\
$^4$ Max-Planck-Institut f\"ur Physik Komplexer Systeme, N\"othnitzer Str. 38, 01187 Dresden, Germany}
\end{center}

We here present details of the renomalization group calculation to the one-loop order using the Euclidean partition function. \cite{justin} For the ease of calculation, we here rewrite the coupling of the order parameter to the massless Dirac fermions, shown in Eq. (4) of the main text as 
\begin{equation}
L_{bf}=g[({\rm Re}\Phi){\bar \Psi}\Psi+({\rm Im}\Phi){\bar \Psi}i\gamma_5\Psi] \: \equiv \:
g(\Phi{\bar\Psi}P_+\Psi+\Phi^*{\bar\Psi}P_-\Psi),
\end{equation}
where the projectors are $P_\pm\equiv\frac{1}{2}(1\pm\gamma_5)$.
 
\section{ Computation of Bosonic self-energy}
\begin{figure}
{\includegraphics[width=16cm,height=2.25cm]{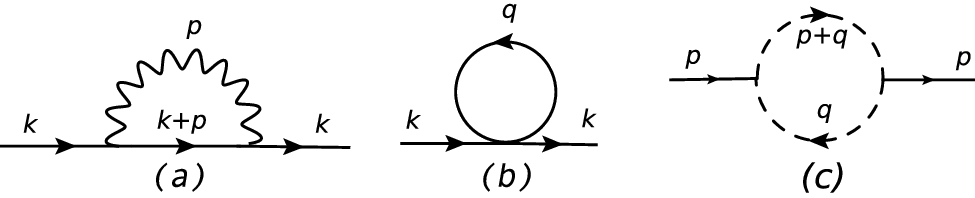}}
 \caption[] {Diagrams contributing to the self energy corrections of the bosonic order parameter field.}
\label{OPself}
\end{figure}

First we consider the renormalization of the self energy of the order parameter field, arising from the diagrams shown in Fig. \ref{OPself}. The one loop correction of the bosonic self energy due to its coupling with the gauge field (see Fig. \ref{OPself}(a)) reads
\begin{equation}
(1a)=-(2 i e)^2 \int \frac{d^d p}{(2 \pi)^d} \: \frac{1}{p^2} \: \left( \delta_{\mu \nu} - \frac{p_\mu p_\nu}{p^2}\right) (2 k+p)_\mu (2 k+p)_\nu \:
\frac{1}{(k+p)^2+m^2} = 4 e^2 \left[ I_1 -I_2 \right],
\end{equation}
where
\begin{equation}
I_1 = \int \frac{d^d p}{(2 \pi)^d} \: \frac{(2 k +p)^2}{p^2 \left[ (k+p)^2 + m^2 \right]} = \int^{1}_0 dx \int \frac{d^d p}{(2 \pi)^d} \:
\frac{(2-x)^2 k^2+p^2}{\left[ p^2 + x(1-x)k^2+ x m^2 \right]^2}
\end{equation}
and
\begin{equation}
I_2= \int \frac{d^d p}{(2 \pi)^d} \: \frac{p_\mu (2 k+p)_\mu p_\nu (2 k+p)_\nu}{p^4 \left[ (k+p)^2 +m^2 \right]}=
\int^{1}_0 dx \: 2 (1-x) \int \frac{d^d p}{(2 \pi)^d} \: \frac{\left( \left( p-x k\right)^2 + 2 k \cdot p - 2 x k^2 \right)^2}{\left[ p^2 + x(1-x) k^2+m^2\right]^3}
\end{equation}
after setting $p+x k \rightarrow p$ in both the integrals. Performing some standard integrals we arrive at
\begin{equation}
I_1=\frac{N_d}{\epsilon} \left( 2 k^2-m^2\right) + {\cal O}(1), \quad I_2=\frac{N_d}{\epsilon} \left( -k^2 -m^2\right)+{\cal O}(1).
\end{equation}
Therefore the correction to the bosonic self energy due to its coupling with the gauge field is
\begin{equation}
(1a)= 12 \: e^2 \: k^2 \: \frac{N_d}{\epsilon} + {\cal O}(1).
\end{equation}
Besides the above renormalization, the bosonic self energy acquires additional correction from the interaction vertex ($\lambda |\Phi|^4$), as shown in Fig. \ref{OPself}(b). This contribution is
\begin{equation}
(1b)= \frac{\lambda}{2} \; \left( 2 N_b + 2 \right) \: \int \; \frac{d^d q}{(2 \pi)^d} \: \frac{1}{q^2 +m^2} \:=\: \lambda \; (N_b+1) \;
\frac{N_d}{\epsilon} \; m^{2-\epsilon} + {\cal O}(1).
\end{equation}
Apart from the above two contributions, additional renormalization of the bosonic self energy comes from its coupling with the massless Dirac fermions, as shown in Fig. \ref{OPself}(c), yielding
\begin{equation}
(1c)=g^2 \; Tr\; \int \; \frac{d^d q}{(2 \pi)^d} \: P_+ \: \frac{\slashed{p}+\slashed{q}}{(p+q)^2} \: P_- \; \frac{\slashed{q}}{q^2} \:=\:
- g^2 \; N_f \; p^2 \; \frac{N_d}{\epsilon} + {\cal O} (1),
\end{equation}
where $\slashed{q}\equiv\gamma^\mu q_\mu$. These three diagrams together give the following renormaization conditions
\begin{eqnarray}
Z_\Phi+ g^2 N_f \frac{1}{\epsilon}- 12 e^2 \frac{1}{\epsilon}=1 \\
Z_\Phi m^2_0 \mu^{-\epsilon}- \lambda \left(N_b+1 \right) m^2 \frac{1}{\epsilon} =m^2,
\end{eqnarray}
leading to $Z_\Phi$ in Eq. (8) and mass renormalization in Eq. (12) in the main text.

\section{ Self energy correction of the gauge fields}
\begin{figure}[t]
{\includegraphics[width=16cm,height=2.25cm]{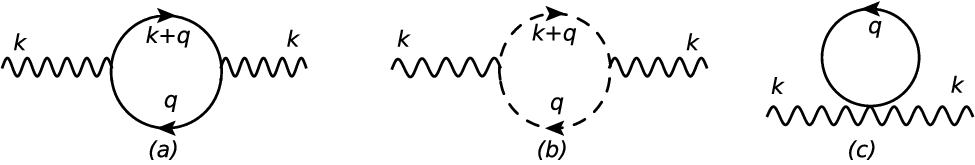}}
 \caption[] {Diagrams contributing to the self energy corrections of the gauge field.}
\label{gaugefieldself}
\end{figure}
Next we present the computation of the self energy correction of the gauge fields. It arises from the diagrams shown in Fig. \ref{gaugefieldself}. Contribution from the diagram (a) is
\begin{equation}
(2a)=N_b \: (2 e)^2 \int \frac{d^d q}{(2 \pi)^d} \: \frac{(k+2 q)_\mu (k+2 q)_\nu}{\left[ q^2 +m^2 \right] \: \left[ (q+k)^2 +m^2 \right]} =
\left( 4 \; N_b \; e^2 \right) \int^1_0 dx \int \frac{d^d q}{(2 \pi)^d} \frac{\frac{4}{d}\; q^2 \delta_{\mu \nu} + (1-2 x)^2 k_\mu k_\nu}{\left[ q^2
+ x (1-x) k^2 +m^2\right]^2},
\end{equation}
once we take $q+ x k \rightarrow q$. Performing some standard integrals gives
\begin{equation}
(2a)= - \; \frac{4}{3} \; N_b \; e^2 \frac{N_d}{\epsilon} k^2 \left( \delta_{\mu \nu} -\frac{k_\mu k_\nu}{k^2} \right) \:-\: 8 N_b e^2 \; \frac{N_d}{\epsilon} \; \delta_{\mu \nu} m^2 + {\cal O} (1).
\end{equation}
Upon taking the contribution from the diagram in Fig. \ref{gaugefieldself}(c), the term proportional to $m^2$ in the last expression exactly cancels out and the gauge field remains transverse. The non-trivial coupling of the gauge field with Dirac fermions also gives a correction to the gauge field propagator, as shown in Fig.\ \ref{gaugefieldself}(b). Contribution from this diagram reads as
\begin{eqnarray}
(2b)=e^2 \; Tr \; \int \frac{d^d q}{(2 \pi)^d} \: \: \frac{\gamma_\mu \slashed{q} \gamma_\nu \left(\slashed{k} + \slashed{q} \right)}{q^2 \; (k+q)^2}
=e^2 \; Tr \left[ \gamma_\mu \gamma_\rho \gamma_\nu \gamma_\sigma \right] \: I_{\rho \sigma} \;=\; e^2 4 N_f \; \left(
\delta_{\mu \rho} \delta_{\nu \sigma}- \delta_{\mu \nu} \delta_{\rho \sigma} + \delta_{\mu \sigma} \delta_{\nu \rho}
\right) \: I_{\rho \sigma},
\end{eqnarray}
where
\begin{equation}
I_{\rho \sigma}= \; \int \frac{d^d q}{(2 \pi)^d} \: \frac{q_\rho (k+q)_\sigma}{q^2 \; (k+q)^2} = \int^1_0 dx \; \int \frac{d^d q}{(2 \pi)^d} \:
\frac{q_\rho q_\sigma - x(1-x) k_\rho k_\sigma}{\left[q^2 + x(1-x) k^2 \right]^2},
\end{equation}
once we take $q + x k \rightarrow q$. A tedious, but otherwise straightforward calculation yields
\begin{equation}
I_{\rho \sigma}\;=\; - \frac{1}{3} \: \left( \frac{\delta_{\rho \sigma}}{2} \;+\; k_\rho k_\sigma \right) \frac{N_d \; k^{-\epsilon}}{\epsilon}
\;+\; {\cal O}(1).
\end{equation}
Therefore the correction to the gauge propagator from the diagram in Fig. \ref{gaugefieldself}(b) is
\begin{equation}
(2 b)=\frac{4}{3} \; N_f e^2 \: k^2 \left[ \delta_{\mu \nu} - \frac{k_\mu k_\nu}{k^2} \right] \: \frac{N_d \; k^{-\epsilon}}{\epsilon} \;+\; {\cal O} (1).
\end{equation}
Collecting the contributions from these three diagrams gives the renormalization condition for $Z_A$ as
\begin{equation}
Z_A + \; e^2 \; \frac{4}{3} \left(N_b \;+\; N_f \right) \frac{N_d}{\epsilon}= 1,
\end{equation}
leading to the announced result in the main text, Eq. (9).

\section{Correction to the self energy of Dirac fermion}
\begin{figure}[t]
{\includegraphics[width=12cm,height=2.25cm]{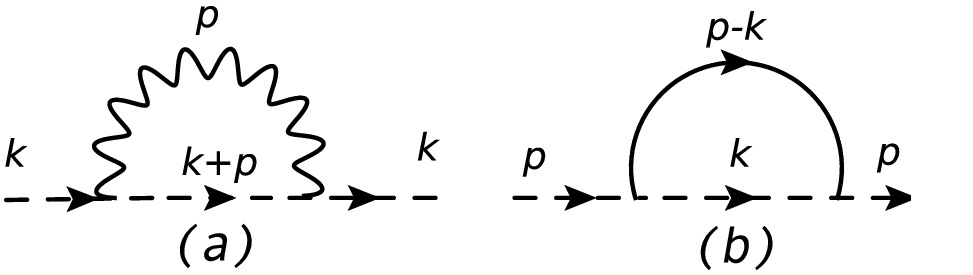}}
 \caption[] {Diagrams contributing to the self energy corrections of the Dirac fermions.}
\label{fermionself}
\end{figure}

Let us now compute the fermionic self-energy corrections to the one loop order. Fermions coupling with the gauge field, shown in Fig. \ref{fermionself}(a), leads to the following correction to its self-energy
\begin{eqnarray}
(3a)=&-& e^2 \int \frac{d^d p}{(2 \pi)^d} \gamma_5 \gamma_\mu \: \frac{i \left(\slashed{p}+\slashed{k} \right)}{(p+k)^2}\: \gamma_5 \gamma_\nu \: \frac{1}{p^2} \: \left( \delta_{\mu \nu} - \frac{p_\mu p_\nu}{p^2}\right) \\
&=&
-i \; e^2 \left[ (2-d) \gamma_\rho \int \frac{d^d p}{(2 \pi)^d} \: \frac{(p+k)_\rho}{p^2 (k+p)^2} -
\gamma_\mu \gamma_\rho \gamma_\nu \int \frac{d^d p}{(2 \pi)^d} \: \frac{p_\mu (k+p)_\rho p_\nu}{p^4 \left(k+p\right)^2}
\right] \\
&=& i \; e^2 \int^1_0 dx \int \frac{d^d p}{(2 \pi)^d} \left[
(2-d) \gamma_\rho \: \frac{p_\rho + (1- x) k_\rho}{\left[ p^2 + x (1-x) k^2 \right]^2} -
(1-x) \gamma_\mu \gamma_\rho \gamma_\nu \; \frac{(p-x k)_\mu (p-x k)_\nu \left( p+(1-x)k \right)_\rho}{\left[p^2+x(1-x)k^2 \right]^3}
\right],
\end{eqnarray}
where $p+ x k \rightarrow p$. After performing the integrals we find a trivial contribution
\begin{equation}
(3a)=- i \; e^2 \; \left[ (2-d) -\frac{1}{3} \left(2 - 2 d \right) \right] \frac{\slashed{k}}{\epsilon} \; k^{-\epsilon} \; \frac{1}{(4 \pi)^2} + {\cal O}(1)\equiv 0 + {\cal O}(1),
\end{equation}
as $d \rightarrow 4$. However, the fermionic self energy acquires nontrivial correction due to its coupling with the order-parameter field, shown in Fig.\ \ref{fermionself}(b), yielding
\begin{equation}
(3b)=g^2 \int \frac{d^d k}{(2 \pi)^d} \: \left[ P_+ \frac{i \slashed{k}}{k^2} P_- +  P_- \frac{i \slashed{k}}{k^2} P_+ \right] \:
\frac{1}{(k-p)^2+m^2} \;=\; g^2 \; \left( \frac{1}{2 \epsilon} \right) \; i \slashed{p} \; N_d m^{-\epsilon} + {\cal O} (1).
\end{equation}
Therefore
\begin{equation}
Z_\Psi=1-\frac{g^2}{2 \epsilon} \; N_d \; m^{-\epsilon},
\end{equation}
as shown in Eq.\ (8) in the main part of the paper.

\section{Renormalization of boson-fermion vertex}
\begin{figure}[t]
{\includegraphics[width=12cm,height=4.25cm]{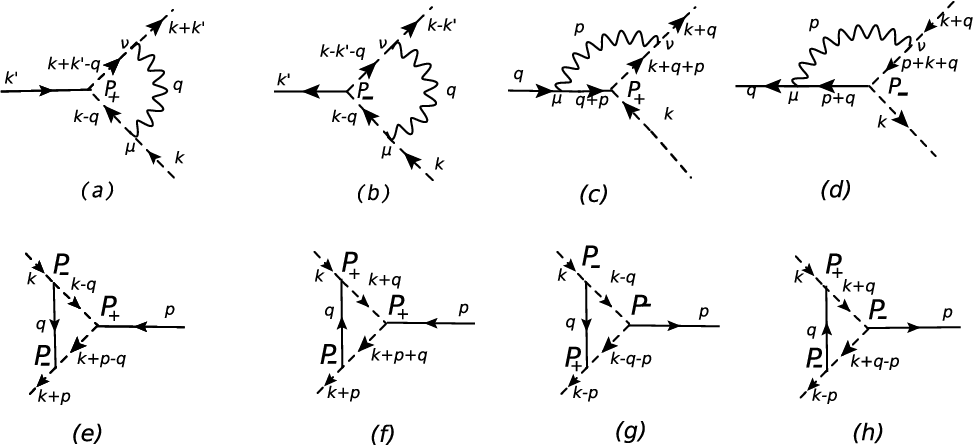}}
 \caption[] {Renormalization of the fermion-boson vertex.}
\label{bosfermvert}
\end{figure}
To the one loop order the boson-fermion vertex also gets renormalized. Its renormalization comes from the diagrams shown in Fig \ref{bosfermvert}. Let us first consider the contribution of the diagram (a), renormalizing $P_+$ vertex
\begin{eqnarray}
(4a)=&-& (i e)^2 \; \int \frac{d^d q}{(2 \pi)^d} \: \frac{1}{ q^2} \: \left( \delta_{\mu \nu} - \frac{q_\mu q_\nu}{q^2}\right) \gamma_5 \gamma_\mu
\frac{\slashed{k}-\slashed{q}}{(k-q)^2} \: P_+ \: \frac{\slashed{k}+\slashed{k}'-\slashed{q}}{(k+k'-q)^2} \; \gamma_5 \gamma_\nu \\
&=& - P_+ \: e^2 \; \int \frac{d^d q}{(2 \pi)^d} \: \bigg[ \gamma_\mu \gamma_\rho \gamma_\sigma \gamma_\mu I_{\rho \sigma}
- \gamma_\mu \gamma_\rho \gamma_\sigma \gamma_\nu I_{\mu \nu \rho \sigma} \bigg],
\end{eqnarray}
where
\begin{eqnarray}
I_{\rho \sigma}= \; \int \frac{d^d q}{(2 \pi)^d} \: \frac{(k-q)_\rho (k+k'-q)_\sigma}{q^2 \; (k-q)^2 \; (k+k'-q)^2}
= \int^1_0 2 y dy \: \int^{1}_0 dx \; \; \int \frac{d^d q}{(2 \pi)^d} \: \frac{\left(q+(1-y)p-(1-xy)k\right)_\rho \left( q+ x y k -y p\right)_\sigma}{\left[ q^2- \left( (1-y)p + x y k\right)^2 + x y k^2 + (1-y) p^2 \right]^3}, \nonumber \\
\end{eqnarray}
and
\begin{eqnarray}
I_{\mu \nu \rho \sigma}&=& \; \int \frac{d^d q}{(2 \pi)^d} \: \frac{(q-k)_\rho (q-p)_\sigma q_\mu q_\nu}{q^4 (q-k)^2 (q-p)^2} \nonumber \\
&=&6 \int^1_0 y^2 dy \int^1_0 (1-x) dx  \int \frac{d^d q}{(2 \pi)^d} \: \left[  \frac{Q_\mu(q,p,k,x,y) Q_\nu(q,p,k,x,y) \left[ q+(1-y)p+(xy-1)k\right]_\rho \left[ q- y p+ xy k \right]_\sigma}{\left[q^2-\left((1-y)p + xy k\right)^2 + (1-y)p^2 + xy k^2 \right]^4} \right], \nonumber \\
\end{eqnarray}
after shifting the momentum as $q-(1-y)p- xyk \rightarrow q$ in $I_{\rho \sigma}$, $I_{\mu \nu \rho \sigma}$, and $Q(q,p,k,x,y)\equiv q+(1-y)p+ xyk $. For our renormalization group calculation, since we use the minimal-subtraction scheme, we only need to keep the divergent pieces from these two integrals, which read
\begin{eqnarray}
I^{div}_{\rho \sigma} &=& N_d \; \delta_{\rho \sigma} \; \frac{1}{4 \epsilon} \; k^{-\epsilon}, \\
I^{div}_{\mu \nu \rho \sigma} &=& 6 \int^1_0 y^2 (1-x) dx dy \int \frac{d^d q}{(2 \pi)^d} \: \frac{q_\mu q_\nu q_\rho q_\sigma}
{\left[ q^2+ (1-y) p^2 + xy k^2 -\left( (1-y)p + xy k \right)^2 \right]} \nonumber \\
&=&\frac{N_d}{24 \; \epsilon} \:k^{-\epsilon} \: \left( \delta_{\mu \nu} \delta_{\rho \sigma} + \delta_{\mu \rho} \delta_{\nu \sigma} + \delta_{\mu \sigma} \delta_{\rho \nu}\right).
\end{eqnarray}
Hence the renormalization of the boson-fermion vertex is
\begin{equation}
(4a)=-  P_+ \; e^2 \: \left[ \gamma_\mu \gamma_\rho \gamma_\sigma \gamma_\mu \delta_{\rho \sigma} \frac{N_d}{4 \epsilon} -
\gamma_\mu \gamma_\rho \gamma_\sigma \gamma_\nu \left( \delta_{\mu \nu} \delta_{\rho \sigma} + \delta_{\mu \rho} \delta_{\nu \sigma} + \delta_{\mu \sigma} \delta_{\rho \nu}\right) \; \frac{N_d}{24 \epsilon}
\right] + {\cal O}(1)= - P_+ \; \frac{3 \; e^2}{ \epsilon} \; N_d + {\cal O}(1).
\label{vertexcorrection}
\end{equation}
The renormalization of the other boson-fermion vertex, proportional to $P_-$, comes from the diagram (b) of Fig. \ref{bosfermvert}, which gives an identical contribution as in Eq. [\ref{vertexcorrection}]. In principle, the boson-fermion vertex can also be renormalized from the diagram Fig. \ref{bosfermvert} (c) and (d). They respectively renormalizes the $P_+$ and $P_-$ vertex. However, it is sufficient to consider one of them, for example (c). Its contribution reads as
\begin{eqnarray}
(4 c)&=&- (i e) \: (2 i e) \: \int \frac{d^d p}{(2 \pi)^d} \: \frac{1}{p^2} \left( \delta_{\mu \nu}- \frac{p_\mu \; p_\nu}{p^2}\right) \; (2 q+p)_\mu
\; \frac{1}{(p+q)^2+m^2} \; P_+ \; \frac{\slashed{q}+\slashed{p}+\slashed{k}}{(q+p+k)^2} \; \gamma_5 \gamma_\nu \nonumber \\
&=& -2  P_+ e^2 \int \frac{d^d p}{(2 \pi)^d} \: \bigg[ \gamma_\alpha \gamma_\nu I_{\nu \alpha} - \gamma_\alpha \gamma_\nu I_{\mu \nu \mu \alpha} \bigg],
\end{eqnarray}
where
\begin{eqnarray}
I_{\nu \alpha}&=&\int \frac{d^d p}{(2 \pi)^d} \: \frac{(2 q+p)_\nu \; \left( q+k'\right)_\alpha}{p^2 \; \left[ (p+q)^2+m^2\right] \;(p+k')^2} \nonumber \\
&=& \int^1_0 2 y \; dy \; \int^1_0 \; dx \; \int \frac{d^d p}{(2 \pi)^d} \frac{\left( p+(1+y)q-xyk \right)_\nu \: \left(p-(1-y)q+(1- xy)k' \right)_\alpha}{\left[ p^2 + \Delta\right]^3},
\end{eqnarray}
where, $k'=k+q$ and $\Delta=xy k'^2 + \left( 1-y \right) \left(q^2+m^2 \right)-\left[ (1-y)q - x y k' \right]^2$. We shift the momentum $p+(1-y)q+xyk' \rightarrow p$, while writing the last equation. The other integral $I_{\mu \nu \mu \alpha}$ reads as
\begin{eqnarray}
I_{\mu \nu \mu \alpha} &=&\int \frac{d^d p}{(2 \pi)^d} \: \frac{p_\mu p_\nu \left( 2 q+p\right)_\mu \left( p+k'\right)_\alpha}{p^4 \; \left[ (p+q)^2+m^2\right] \;(q+k')^2} \nonumber \\
&=& 6 \int^1_0 y^2 \; dy \int^1_0 (1-x) \; dx \: \frac{H_\mu(x,y,q,p,k') H_\nu(x,y,q,p,k') \left( p+(1+y)q-xy k' \right)_\mu
\left( p-(1-y)q-(xy-1) k' \right)_\alpha}{\left[ p^2 +\Delta \right]^4 }, \nonumber \\
\end{eqnarray}
after shifting the momentum $p$ as before, where $H(x,y,q,p,k') \equiv p-(1-y)q - x y k'$. We are however interested only in the divergent pieces of these two integrals which read as
\begin{eqnarray}
I^{div}_{\nu \alpha}&=& \frac{N_d}{ 4 \; \epsilon } \: k^{-\epsilon} \delta_{\nu \alpha}, \\
I^{div}_{\mu \nu \mu \alpha} &=& \frac{N_d}{ 2 4 \; \epsilon } \; k^{- \epsilon} \: \bigg( \delta_{\mu \nu} \delta_{\mu \alpha}+ \delta_{\mu \mu} \delta_{\alpha \nu}+ \delta_{\mu \alpha} \delta_{\nu \mu} \bigg).
\end{eqnarray}
Hence the renormalization of the $P_+$ vertex from the diagram (c) in Fig. \ref{bosfermvert} reads as
\begin{equation}
(4 c)= -2 \; P_+ \; e^2 \bigg[ \frac{\gamma_\alpha \gamma_\nu \; \delta_{\alpha \nu}}{4} -\frac{\gamma_\alpha \gamma_\nu}{24}
\left(2 \delta_{\mu \alpha} \delta_{\mu \nu} + \delta_{\mu \mu} \delta_{\alpha \nu} \right)\bigg] \: \frac{N_d}{\epsilon} + {\cal O} (1)
=0 + {\cal O} (1),
\end{equation}
as $d \rightarrow 4$. Hence this diagram provides trivial renormalization of the boson-fermion vertex ($P_+$). Similarly, one can show that renormalization of the other boson-fermion vertex ($P_-$) arising from the diagram Fig. \ref{bosfermvert} (d) is also trivial. Renormalization of the Yukawa vertex can also arise from boson-fermion interaction, as shown in Fig. \ref{bosfermvert} (e), (f), (g), (h). The combined contribution of these diagram is however
\begin{eqnarray}
(4 e)+ (4 f) +(4 g) + (4 h)&=& g^2 \int \frac{d^d q}{(2 \pi)^{d}} \bigg\{ 
\bigg( P_+ \: \frac{i \left( \slashed{k}+\slashed{p}-\slashed{q} \right)}{\left(k+p-q \right)^2} \: P_+ \: \frac{i \left( \slashed{k}-\slashed{q} \right)}{\left(k-q \right)^2} \: P_-\bigg)
+ \bigg( P_- \: \frac{i \left( \slashed{k}+\slashed{p}+\slashed{q} \right)}{\left(k+p+q \right)^2} \: P_+ \: \frac{i \left( \slashed{k}+\slashed{q} \right)}{\left(k+q \right)^2} \: P_+\bigg) \nonumber \\
&+& \bigg( P_+ \: \frac{i \left( \slashed{k}-\slashed{p}-\slashed{q} \right)}{\left(k-p-q \right)^2} \: P_- \: \frac{i \left( \slashed{k}-\slashed{q} \right)}{\left(k-q \right)^2} \: P_-\bigg)+
\bigg( P_- \: \frac{i \left( \slashed{k}-\slashed{p}+\slashed{q} \right)}{\left(k-p+q \right)^2} \: P_- \: \frac{i \left( \slashed{k}+\slashed{q} \right)}{\left(k+q \right)^2} \: P_+\bigg)
\bigg\} \: \frac{1}{q^2+m^2} \nonumber \\
& \equiv & \: 0.
\end{eqnarray}

These contributions yield the announced renormalization condition of the Yukawa coupling (g) in the main text, namely
\begin{equation}
Z_\psi \; Z_\phi^{1/2} g_0 \; \mu^{-\epsilon} + 3 e^2 g \; \frac{1}{\epsilon} =g.
\end{equation}

\section{Renormalization of $|\Phi|^4$ vetrex}
\begin{figure}[t]
{\includegraphics[width=10cm,height=2.25cm]{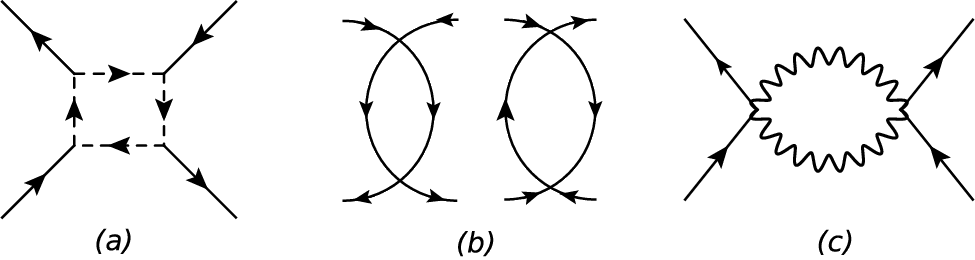}}
 \caption[] {Renormalization of the interaction vertex of boson.}
\label{lambdvert}
\end{figure}
Finally we consider the renormalization of the $|\Phi|^4$ vertex. Renormalization of this vertex from the Yukawa coupling, shown in Fig. \ref{lambdvert}(a), reads
\begin{equation}
(5a)=-2 \; g^4 \;Tr \; \int \; \frac{d^d q}{(2 \pi)^d} \: \frac{P_+ \; \slashed{q}\; P_- \left( \slashed{p}_1 +\slashed{q} \right) \; P_+\;
\left( \slashed{p}_1 + \slashed{q}-\slashed{p}_4 \right) \; P_- \; \left( \slashed{p}_1 + \slashed{q}-\slashed{p}_3-\slashed{p}_4\right)}
{q^2 \; (p_1+q)^2 \; (p_1+q-p_4)^2 \; (p_1+q-p_3-p_4)^2} = - 8 g^4 N_f \: \frac{N_d}{\epsilon} + {\cal O}(1).
\end{equation}
The renormalization of this vertex arising from the diagrams in Fig. \ref{lambdvert}(b) is
\begin{equation}
(5b)=\lambda^2 \; \left( N_b + 4 \right) \: \frac{N_d}{\epsilon} \;+\; {\cal O} (1).
\end{equation}
Moreover, the coupling of the order parameter field with the gauge field also provides a renormalization of the $|\Phi|^4$ vertex, as shown in Fig. \ref{lambdvert}(c). Its contribution reads
\begin{eqnarray}
(5c)=\; -\; 32 \; e^4 \; \int \; \frac{d^d q}{(2 \pi)^d} \: \frac{1}{q^4} \: \left[ \delta_{\mu \nu} \;-\; \frac{q_\mu q_\nu}{q^2}\right] \:
\left[ \delta_{\mu \nu} \;-\; \frac{q_\mu q_\nu}{q^2}\right] \;=\; -32 \; e^4  \; \frac{3 N_d}{\epsilon}.
\end{eqnarray}
Upon collecting contributions from all the diagrams in Fig. \ref{lambdvert}, one gets the renormalization condition of the coupling constant $\lambda$, as in Eq. (11) in the main part of the paper.


\begin{thebibliography}{99}
\bibitem{gamma-insulator} Here the gamma matrices are defined as $\gamma_0=I_2 \otimes \sigma_3, \gamma_1=\sigma_3 \otimes \sigma_2, \gamma_2=I_2 \otimes \sigma_2, \gamma_3=\sigma_1 \otimes \sigma_2$, and $\gamma_5=\sigma_2 \otimes \sigma_2$.
\end{thebibliography}
\end{document}